\documentstyle[aps,twocolumn]{revtex}
\draft
\begin{document}
\author{Yan V. Fyodorov$^{}$\cite{leave}  and Hans-J\"{u}rgen Sommers}
\address{ Fachbereich Physik,
Universit\"{a}t-Gesamthochschule Essen\\
 Essen 45117, Germany}
\title{Statistics of S-matrix poles in few-channel chaotic scattering:\\
crossover from isolated to overlapping resonances}

\date{July,26 1995}
\maketitle

\begin{abstract}
We derive the explicit expression for the distribution of resonance
widths in a chaotic quantum system coupled to continua via M
equivalent open channels. It describes a crossover from the $\chi^2$
distribution (regime of isolated resonances) to a broad power-like
distribution typical for the regime of overlapping resonances. The
first moment is found to reproduce
exactly the Moldauer-Simonius relation between the mean resonance
width and the transmission coefficient. This fact may serve as
another manifestation of equivalence between the spectral and the
ensemble averaging.
\end{abstract}

\narrowtext

Chaotic scattering has been a subject of rather intensive research
activity during the last decade, both theoretically and
experimentally, see reviews \cite{Gasp}-\cite{ion}. This phenomenon
is encountered in a variety
of physical systems ranging from atoms, atomic nuclei and molecules
to mesoscopic devices \cite{Stone} and microwave cavities
\cite{cav,Richter}.

One of the basic concepts in the domain of chaotic quantum
scattering is the notion of resonances, representing long-lived
intermediate states to which bound states of a "closed" system are
converted due to coupling to continua. On a formal level resonances
show up as poles of the scattering matrix $S_{ab}(E)$ occuring at
complex energies $ E_{k}={\cal E}_{k}-\frac{i}{2}\Gamma_{k}$, where
${\cal E}_{k}$ and $\Gamma_{k}$ are called position and width of the
resonance,
correspondingly. The general problem of determining the domain of
concentration and distribution of S-matrix poles is of fundamental
interest in the scattering theory and  is a field of intensive
research activity
\cite{Zworski}. Powerful numerical methods are available ( e.g  the
method of complex scaling) allowing one to extract resonance
parameters for models in atomic and molecular physics\cite{Kukulin}.

Whereas the issue of energy level statistics in closed chaotic
systems was addressed in an enormous amount of papers (see,e.g.
\cite{Bohigas} and references therein), statistical characteristics
of resonances are much less studied and attracted significant
attention only recently. In the case of weak effective coupling to
continua individual resonances do not overlap:
$\langle\Gamma\rangle\ll \Delta$, with $\Delta$ standing for the
mean level spacing of the
"closed" system and $\langle\Gamma\rangle$ standing for the mean
resonance width. Under these conditions one can use a simple first
order perturbation theory to calculate resonance widths in terms of
eigenfunctions of the closed system, see e.g \cite{Stone,Altsh}.
Quite generally, one finds in such a procedure that the scaled
widths $y_{s}=\frac{\Gamma}{\langle\Gamma\rangle}$ are distributed
according to the so-called $\chi^2$-distribution:
\begin{equation}\label{chi}
\rho(y_{s})=\frac{(\nu/2)^{\nu/2}}{\Gamma(\nu/2)}y_{s}^{\nu/2-1}
e^{-\frac{\nu}{2} y_{s}}
\end{equation}
 where $\Gamma(z)$ stands for the Gamma function and
the parameter $\nu=M $ $(\nu=2M)$ for systems with preserved
(broken) time reversal symmetry (TRS), $M$ being the number of open
scattering channels. The basic fact underlying this simple form of
eq.
(\ref{chi}) is the Gaussian statistics of components of wavefunctions.
 The case $\nu=1$ is known as Porter-Thomas
distribution\cite{Porter}. It was shown to be in agreement with
experimental data in neutron-nuclei resonances, the fluorescence
excitation spectrum of the $NO_2$ molecule, the diamagnetic Rydberg
spectrum of the lithium atom (see references in
\cite{Gasp},\cite{Del},\cite{Porter}) and, more recently in
microwave
cavities \cite{Richter}.

When the coupling to continua increases some resonances start to
overlap and the simple perturbative result eq.(\ref{chi}) loses its
validity. This regime is most easily reached when the number of open
channels M is of the same order as the number $N$  of bound states
in the closed system, the situation encountered in the so-called
"three disk" scattering problem \cite{Gasp}. The range of parameters
$M\propto N$ was studied in much detail in \cite{Som}. It
corresponds to very strong overlap of resonances
$\langle\Gamma\rangle>>\Delta$ which is the opposite limiting case
as compared to the domain of validity of the distribution
eq.(\ref{chi}).

In the general case of M open channels, $M$ very unstable states
(broad resonances) are formed, whereas the rest, $N-M$ states, have
much smaller widths \cite{Sok,Frank}. Changing the effective
coupling to continua one arrives at the situation when a substantial
portion of these long-lived resonances
show a considerable overlap, even for the case of few open channels
$M\ll N$.
Experimentally, one quite frequently encounter the case of only
$M\sim 1$ open channels  and  $\langle \Gamma\rangle\sim \Delta$
\cite{ion,Stone,Richter,Del,Schinke}. In this situation neither
application of eq.(\ref{chi}) nor exploitation of the results from
\cite{Som} can be reasonably justified. The main goal of the
present publication is to fill in this gap providing the explicit
distribution covering the whole range of the parameter
$\langle \Gamma\rangle/ \Delta$ for the few channel scattering problem.

At present, there are two complementary theoretical tools employed
to calculate statistical characteristics of "open" quantum systems
whose closed classical counterparts demonstrate chaotic behaviour.
These are the semiclassical approach \cite{Gasp,Smilansky} and the
stochastic approach \cite{Weid},
the relation between  both methods being comprehensively discussed
in \cite{Lew}. In particular, the semiclassical approach operates
with genuine microscopic Hamiltonians and allows for treating
particular systems with full account of their specific features.
However, the applicability
of the method is essentially restricted to the case of many open
channels $M\gg 1$. In contrast, the stochastic approach could, in
principle, deal with as many open channels as one likes. As a price,
it is able to address only {\it universal}, i.e. system independent
characteristics common to most of quantum
chaotic systems and stemming from the very basic fact of chaotic
internal motion. Such kind of generality is achieved by
incorporating the random matrix description of chaotic systems as an
input information. The latter principle is a commonly accepted one
in the domain of Quantum Chaos \cite{Bohigas}.

The starting point of the stochastic approach is the following
multichannel Breit-Wigner representation for elements of the
scattering matrix \cite{Weid}:
\begin{equation}\label{def}\begin{array}{c}
\displaystyle{S_{ab}(E)=\delta_{ab}-2i\pi\sum_{ij}
W_{ia}^{*}[E-{\cal H}_{ef}]_{ij}^{-1}W_{jb}}\\
\left({\cal H}_{ef}\right)_{ij}=H_{ij}-i\pi\sum_{a}W_{ia}W_{ja}^{*}
\end{array}\end{equation}
Here the $N\times N$ matrix $H_{ij}$ is to model the Hamiltonian of
the closed chaotic system and thus is chosen to be a member of the
corresponding Gaussian ensemble of random matrices. To be specific,
we consider in the present publication only systems with broken TRS
by taking $H_{ij}$ to be a random Hermitian matrix from the Gaussian
Unitary Ensemble \cite{Bohigas}. The amplitudes $W_{ai},\quad
a=1,2,...,M$ are matrix elements coupling the internal motion to one
out of M open
channels. Without much loss of generality these amplitudes can be
chosen in a way ensuring that the average
$S-$matrix is diagonal in the channel basis:
$\overline{S_{ab}}=\delta_{ab}\overline{S_{aa}}$.  Provided the
energy $E$ is real, one finds the following expression:\cite{Weid}:
\begin{equation}\label{sav}
\overline{S_{aa}}=\frac{1-\gamma_{c}g(E)}{1+\gamma_{c}g(E)};\quad
\gamma_{c}=\pi\sum_{i}W^{*}_{ia}W_{ia}
\end{equation}
where $g(E)=i E/2+(1-E^2/4)^{1/2}$ and we assumed that all $M$
channels are statistically equivalent for the sake of simplicity.
The strength of coupling to continua is convenient to characterize
via the transmission coefficients  $T_{a}=1-|\overline{S_{aa}}|^2$
that are given for the present case by the following expression:
\begin{equation}\label{trans}
T_{a}^{-1}=\frac{1}{2}
\left[1+\frac{1}{2 \mbox{Re}\, g(E)}(\gamma_{c}+\gamma_c^{-1})\right]
\end{equation}
 The quantity  $T_{a}$ measures the part of the  flux in channel
$a$ that spends substantial part of the time in the interaction
region\cite{Weid,Lew}.
We see that both limits
$\gamma_{c}\to 0$ and $\gamma_{c}\to \infty$ equally correspond to
the weak effective coupling regime $T_{a}\ll 1$ whereas the
strongest coupling (at fixed energy $E$ ) corresponds to the value
$\gamma_{c}=1$. The maximal possible coupling corresponding to the
upper bound  $T_a=1$ is achieved in the present model for an energy
interval in the vicinity of the center $E=0$.

In order to get access to the distribution of resonance widths we
use the fact that resonances are actually eigenvalues of the
non-Hermitian Hamiltonian
matrix ${\cal H}_{ef}$ defined in eq.(\ref{def} ).  Two-dimensional
density of these eigenvalues $\rho(X,Y)$ in the complex plane
$E=X+iY$ can be found if one knows the "potential" \cite{Som}:
$$
\Phi(X,Y,\kappa)=\overline{\frac{1}{2\pi N}\ln{Det[(E-{\cal
H}_{ef})(E-{\cal H}_{ef})^{\dagger}+\kappa^2]}}
$$
in view of the relation:$\rho(X,Y)=
\lim_{\kappa\to 0}\partial^2 \,\Phi(X,Y,\kappa)$, where
$\partial^2$ stands for the two-dimensional Laplacian. Technically,
it turns out to be much easier to restore the potential from its
derivative:
\begin{equation}\label{dens}\begin{array}{c}
\displaystyle{\frac{\partial^2 \Phi}{\partial \kappa^2}
=\frac{1}{2\pi N}\frac{d}{d\kappa}
\lim_{\kappa_{b}\to\kappa}\frac{\partial}{\partial \kappa}
\ln{Z(\kappa_b,\kappa)}} \\
Z(\kappa_b,\kappa)=\frac{Det[(E-{\cal H}_{ef})(E-{\cal
H}_{ef})^{\dagger}+\kappa^2]}{Det[(E-{\cal H}_{ef})(E-{\cal
H}_{ef})^{\dagger}+\kappa_{b}^2]}
\end{array}\end{equation}
Then one can find the following representation for the generating
function $Z(\kappa_b,\kappa)$ in terms of the Gaussian integral over
both commuting and anticommuting (Grassmann) variables:
\begin{equation}\label{gau}\begin{array}{c}
(-1)^N Z(\kappa_b,\kappa)=\int[d\Psi]\exp\{-{\cal L}_0 (\Psi)-{\cal
L}_1 (\Psi)\}\\
{\cal L}_0 (\Psi)=
\kappa_{b}(\Psi^{\dagger}\hat{\Lambda}\hat{L}\Psi)+iX(\Psi^{\dagger}
\hat{L}\Psi)-i\Psi^{\dagger}(H\otimes\hat{L})\Psi\\
{\cal L}_1 (\Psi)=-Y(\Psi^{\dagger}\hat{\sigma}_{0}\Psi)-\Psi^{\dagger}
(\hat{\Gamma}\otimes\hat{\sigma}_{0})\Psi+(\kappa-\kappa_b)
(\Psi^{\dagger}\hat{K}\Psi)
\end{array}\end{equation}
where $\Psi^{\dagger}=(\vec{S}^{\dagger}_{1},\vec{S}^{\dagger}_2,
\chi^{\dagger}_1,\chi^{\dagger}_2);\quad
[d\Psi]=\prod_{p=1,2}{d\vec{S}_p d\vec{S}^{\dagger}_{p}
d\chi_1 d\chi^{\dagger}_1}$, with $\vec{S}_p$ and $\chi_p$ being
$N-$component vectors of complex commuting and Grassmannian
variables, respectively.
The $4\times 4$ matrices $\hat{\Lambda},\hat{L},\hat{\sigma}_0$ and
$\hat{K}$
are (block)diagonal of the following structure:
$$ \begin{array}{c}
\hat{\Lambda}=\mbox{diag}(1,-1,1,-1);\quad
\hat{L}=\mbox{diag}(1,-1,1,1)\\
\hat{K}=\mbox{diag}(0,0,1,-1);\quad
\hat{\sigma_0}=\mbox{diag}(i\Sigma_x,\Sigma_x)\end{array}$$
and $\Sigma_x=\left(\begin{array}{cc}0&1\\1&0\end{array}\right)$.
The entries of the $N\times N$ matrix $\hat{\Gamma}$ are given by
$\Gamma_{ij}=
-\pi\sum_{a}W_{ia}W_{ja}^{*}$.

It is easy to notice that if one suppresses the second term ${\cal
L}_1(\Psi)$
in the exponent of eq.(\ref{gau}), the corresponding Gaussian
(super)integral
just coincides with the generating function emerging in
calculations of the two-level
spectral correlation function for the pure GUE \cite{susy}. This
fact allows one to apply the well-developed technique pioneered by
Efetov \cite{susy}
in order to perform the averaging over the GUE of matrices $H$.
After a set of standard manipulations\cite{Weid,susy} one arrives
at the following expression for the density $\rho(y)$ of level
widths $y=\frac{\pi\Gamma}{\Delta}=-2\pi\overline{\rho}(X)N Y$ of
those resonances, whose positions are within
the narrow window around the point $X$ of the spectrum\cite{note2}:
\begin{equation}\label{full}
\rho(y)=\frac{1}{2}\frac{\partial^2}{\partial y^2}\int_{0}^{\infty}
dz z \phi(y,z)\end{equation} where
\begin{equation}
\begin{array}{c}
\phi(y,z)=\\ i \displaystyle{\frac{\partial}{\partial z}}
\int d\mu(Q)Str(\hat{K}\hat{Q}) \, Sdet^{-M/2}\left[1-\frac{i}{2\tau}
(\hat{Q}\hat{\sigma}+\hat{\sigma}\hat{Q})\right]\\
\times\exp\{-\frac{iz}{2}Str(\hat{Q}\hat{\Lambda})-\frac{iy}{2}
Str(\hat{Q}\hat{\sigma})\}\end{array}\end{equation}
where $\hat{\sigma}=\mbox{diag}(\Sigma_x,\Sigma_x)$ and the
(graded) matrices
$\hat{Q}$ satisfying  $\hat{Q}^2=-1$ are taken from the graded
coset space $U(1,1/2)/U(1/1)\otimes U(1/,1)$,
whose explicit parametrization
can be found in \cite{susy}. Here we introduce the notation
$\tau=\frac{1}{2\pi\overline{\rho}(X)}(\gamma_{c}+\gamma_{c}^{-1})$
 and used the symbols $Str,Sdet$ for the graded trace and the
graded determinant, correspondingly.

 Now it is evident, that the form of the resonance widths
distribution does not depend on
$\overline{S_{ab}}$ itself. Rather, it is uniquely
determined by the value of the transmission coefficient equal to
$T_{a}\equiv2/(1+\tau)$ for the present case. This is very
satisfactory feature shared by all observables like crossections,
etc. and believed to be an indicator of universality of the obtained
results\cite{Weid,Lew}.

 Still, the evaluation of the superintegral in eq.(\ref{full}) and
subsequent restoration of the density $\rho(y)$ is quite an
elaborate task. For this reason we  choose to skip all technical
details in favour of presenting the outcoming result, which turns
out to be surprisingly simple:
\begin{equation}
\label{main}\begin{array}{c}
\rho(y=\frac{\pi\Gamma}{\Delta})=\frac{1}{2\Gamma(M)y^2}\int_{y(\tau-1)}^{y(\tau+1)}
dt t^M e^{-t}\\
=(-1)^M\frac{y^{M-1}}{\Gamma(M)}\frac{d^M}{dy^M}
\left(e^{-y\tau}\frac{\sinh{y}}{y}\right)
\end{array}
\end{equation}
This expression provides the explicit form for the distribution of
level widths
for a M-channel chaotic system with broken time-reversal invariance.
It constitutes the main result of the present publication.

Quick inspection of eq.(\ref{main}) shows that it is indeed reduced
to the $\chi^2$
distribution, eq.(\ref{chi}) when the effective coupling to
continua is weak: $\tau\gg 1$. Under this condition resonances are
typically too narrow to overlap with others: $y\lesssim
\tau^{-1}\ll1$.
However, as long as the effective coupling
becomes stronger, the parameter $\tau$ decreases towards unity.
Under these conditions another domain of resonance widths becomes
more and more important: $\frac{M}{\tau+1}< y < \frac{M}{\tau-1}$,
where the distribution eq.(\ref{main}) shows the powerlaw decrease:
$\rho(y)\approx \frac{1}{2}My^{-2}$.
The most drastic difference from eq.(\ref{chi}) occurs for the
maximal effective coupling
$\tau=1$ (i.e $\gamma_c=1$ and $E=0$). In this regime the powerlaw tail
extends up to infinity making all positive moments ( starting from
the first one) to be apparently divergent \cite{com}.

It is interesting to
note that the behaviour $\rho(y)\approx \frac{1}{2}My^{-2}$ nicely
matches that obtained in\cite{Som} for the asymptotic limit
$M\propto N\gg 1$. Another point which is worth to be mentioned is
that evaluating the first moment of the distribution
eq.(\ref{main}) exactly we arrive at the following expression for
the mean resonance widths:
\begin{equation}\label{Simonius}
\frac{\langle\Gamma\rangle}{\Delta}=-\frac{M}{2\pi}\ln{\frac{\tau-1}{\tau+1}}
\equiv -\frac{M}{2\pi}\ln{(1-T_{a})}
\end{equation}
This formula is well known in nuclear physics as Moldauer-Simonius
relation\cite{Mol}. It was derived for systems with preserved TRS by
averaging the $S-$matrix over the energy spectrum and using the
unitarity condition.
The fact that the ensemble averaging produces exactly the same
expression is an explicit manifestation of the validity of the
ergodicity hypothesis and can be considered as another support for
the applicability of the stochastic approach.

Generally, we see, that the logarithmic divergency of $\langle
\Gamma\rangle$ at the critical coupling $\tau=1$ is a direct
consequence of the $1/y^2$
decrease of the probability distribution eq.(\ref{main}). Thus, one
may say, that such a powerlaw tail is typical
for chaotic systems strongly coupled to continua and is just one of the
clear manifestations of the strong overlap between individual resonances.

To this end, it is interesting to mention that a little different
powerlaw distribution of resonance widths - that of the form
$\rho(y)\propto y^{-3/2}$ -
was observed in numerical studies of the "diffusive " quantum
chaotic system
strongly coupled to continua \cite{Shep}. The authors suggested a
transparent qualitative explanation of this effect based on the very
fact of classical
diffusive dynamics.
This argumentation, when appropriately modified, allows one to
understand the form of the powerlaw tail in our case as well.
Indeed, it is well-known that
 the choice of GUE as a model for the closed chaotic system
precludes the diffusive spread  of wavepackets from being taken into
consideration. Instead, the corresponding "chaotic" classical
dynamics can be
visualised as that of a particle hopping in a phase space of $N$
states. The one-step transition probabilities between any pair of
these states are equal.
Among the totality of N states there are M states strongly coupled
to continua:
the particle is "absorbed" (escapes from the sample) when it hits
any of these
M states. Assuming $M/N\ll 1$ the probability to survive up to a
time $\tau$
is simply $p(\tau)\propto \exp{[-\frac{M}{N}\tau]}$. Then the
distribution of "inverse escape time" $\Gamma=\tau^{-1}$ can be
estimated as $\rho(\Gamma)=
\frac{dp(\Gamma^{-1})}{d\Gamma}$ which immediately gives
$\rho(y)\propto M/y^{2}$
for the normalized width $y=N\Gamma$, provided $y\gg M$.

The following comment is appropriate here. Dealing with realistic
models of chaotic systems containing no random parameters one always
performs statistics
over an energy interval $\delta E$ containing many levels: $\delta
E \gg \Delta$, but being small enough for a systematic variation of
the smoothed level density $\overline{\rho(E)}$ to be neglected:
$\delta E\ll \overline{\rho}/
\frac{d \overline{\rho}}{d E}$. One may expect that {\it universal}
features of such a statistics are adequately reproduced within the
framework of the stochastic approach, but  only on the level of
"local-in-spectrum"
characteristics calculated at {\it fixed} value of E. Indeed, any
spectral
averaging in the stochastic model performed on a scale comparable
with the radius of the semicircle unavoidingly mixes up data
corresponding to very different values of the transmission
coefficients, the procedure washing out
any relevant physical information. In particular, it seems quite
meaningless
to consider quantities like the "globally" averaged  resonance
width $\Gamma_{gl}=\frac{1}{N}\sum_{k}\Gamma_k$ which can be
trivially found from the sum rule:
$\Gamma_{gl}/\Delta=\frac{2}{N\Delta}\mbox{Tr Im}{\cal
H}_{ef}=2M\gamma_{c}$. This quantity can not be related to any
particular
transmission coefficient. This fact, however, should not be
misinterpreted as impossibility to have universal statistics of
$S-$matrix poles within
the stochastic approach\cite{Lew}.  Rather, the quantity
$\Gamma_{gl}$ can be found via the direct integration of the {\it
universal} expression $\langle\Gamma\rangle/\Delta$ ,
eq.(\ref{Simonius}), over the energy $E$, upon  substituting there
the energy dependent value $T_a(E)$ from eq.(\ref{trans}).
The result of the integration is finite for any $\gamma_c$, thus
concealing
a specific role of the critical coupling $\gamma_{c}=1$ when
resonances with
a  {\it divergent} local mean width occur sufficiently close to the
center of the spectrum\cite{com}

The best candidates for checking the applicability of eq.(\ref{main})
to real physical systems are realistic models of ballistic
mesoscopic devices
subject to an applied magnetic field that serves to break the
TRS\cite{Stone}.
Very recently there were also reports on realizing TRS breaking in
microwave cavities \cite{TRS}. However, it is not sufficiently clear
whether the specific ways to violate TRS used in the cited papers
can be simulated by simple substitution of
the GUE matrices for $H$ in the framework of the stochastic approach.

The best experimentally accessible resonance data \cite{Richter} as
well as models used to study
chaotic scattering correspond to preserved TRS. In order to treat
this situation in the framework of the stochastic approach one has
to use GOE rather than GUE matrices. All basic steps leading to
eq.(\ref{full}) remain essentially the same, but the graded matrices
$\hat{Q}$ acquire larger size
making the analytical computations very involved.
It is however quite clear that all the basic qualitative features of the
distribution eq.(\ref{main}) (in particular, the powerlaw behaviour
$\rho(y)\propto My^{-2}$ for the overlapping resonance regime)
should show up in that case as well.
This fact is already indicated by the validity of the Moldauer-Simonius
relation for such systems. Moreover, this relation gives support to
 a conjecture, that the expression eq.(\ref{main}) may well turn
out to be
a good approximation
for $M-$channel systems with preserved TRS provided one substitutes
$M/2$ for $M$ into the first form of the eq.(\ref{main}) and interprete
$y$ as $y=\frac{\pi\Gamma}{2\Delta}$.

Authors are very obliged to P.Seba for providing them with his
unpublished numerical data \cite{Seba} suporting the validity of the
distribution eq. (\ref{main}) and to V.Sokolov for numerous
illuminating discussions and critical reading of the manuscript.
Y.V.F. appreciates discussions on chaotic scattering with F.Haake,
N.Lehmann, P.Seba,  H.-J.St\"{o}ckmann and J.Zakrzewski and is
grateful to D.Shepelyansky for attracting his attention to the
powerlaw
distribution found in \cite{Shep}, and to A.Mirlin and B.Khoruzhenko for
constructive remarks.

The financial support by SFB 237 "Unordnung und grosse Fluctuationen"
is acknowledged with thanks.


\begin{thebibliography}{99}
\bibitem{note2} Resonances lie in the lower half plane, hence $Y\le
0$. Widths are normalized to the
mean level spacing of the closed system (in our case,GUE) which is
$\Delta(X)=(\overline{\rho}N)^{-1}$, with $
\overline{\rho}=\frac{1}{2\pi}(4-X^2)^{1/2})$.
\bibitem{com} Formally speaking, for any finite $N$ the widths can
not exceed the upper bound, dictated by the total width $\mbox{Tr
Im}{\cal H}_{ef}$. That means, that the distribution eq.(\ref{main})
is cut at $y\sim N$ and may be also modified
in the small domain of the width $O(1/N)$ around the critical
coupling $\gamma=1$, see F.M.Izrailev et al.{\it Phys.Rev.E} v.49
(1994), 130
\end{thebibliography}
\end{document}